\documentclass[12pt,preprint]{aastex}

\usepackage{amsmath}
\usepackage{amsmath}
\usepackage{amssymb}
\usepackage{amstext}
\usepackage{longtable}
\usepackage{ulem}
\usepackage{graphicx}
\usepackage{epstopdf}
\usepackage{epsfig}
\usepackage{multirow}

\usepackage{color}
\definecolor{myColor}{rgb}{0.9,0.9,0.9}  

\begin{document}
\shorttitle{Nebular water depletion as the cause of Jupiter's low oxygen abundance}
\shortauthors{O. Mousis et al.}
\title{Nebular water depletion as the cause of Jupiter's low oxygen abundance \\
}


 \author{Olivier~Mousis\altaffilmark{1}, Jonathan~I.~Lunine\altaffilmark{2}, Nikku~Madhusudhan\altaffilmark{3}, Torrence~V.~Johnson\altaffilmark{4}}
\altaffiltext{1}{Universit{\'e} de Franche-Comt{\'e}, Institut UTINAM, CNRS/INSU, UMR 6213, Observatoire des Sciences de l'Univers de Besancon, France {\tt olivier.mousis@obs-besancon.fr}}
\altaffiltext{2}{Center for Radiophysics and Space Research, Space Sciences Building Cornell University,  Ithaca, NY 14853, USA}
\altaffiltext{3}{Yale Center for Astronomy and Astrophysics, Department of Physics, Yale University, New Haven, CT 06511}
\altaffiltext{4}{Jet Propulsion Laboratory, California Institute of Technology, 4800 Oak Grove Drive, Pasadena, CA, 91109, USA}

\begin{abstract}
Motivated by recent spectroscopic observations suggesting that atmospheres of some extrasolar giant-planets are carbon-rich, i.e. carbon/oxygen ratio (C/O)~$\ge$~1, we find that the whole set of compositional data for Jupiter is consistent with the hypothesis that it be a carbon-rich giant planet.  We show that the formation of Jupiter in the cold outer part of an oxygen-depleted disk (C/O $\sim$1) reproduces the measured Jovian elemental abundances { at least} as well as the hitherto canonical model of Jupiter formed in a disk of solar composition (C/O = 0.54). The resulting O abundance in Jupiter's envelope is then moderately enriched by a factor of $\sim$2 $\times$ solar (instead of $\sim$7 $\times$ solar) and is found to be consistent with values predicted by thermochemical models of the atmosphere. That Jupiter formed in a disk with C/O $\sim$1 implies that water ice was heterogeneously distributed over several AU beyond the snow line in the primordial nebula and that the fraction of water contained in icy planetesimals was a strong function of their formation location and time. The Jovian oxygen abundance to be measured by NASA's Juno mission en route to Jupiter will provide a direct and strict test of our predictions.
\end{abstract}



\section{Introduction}

Observations of extrasolar planets have revealed the possible existence of a new class of giant planets, the so-called carbon-rich planets (CRPs) (Madhusudhan et al. 2011a). A CRP is defined as a planet with a carbon-to-oxygen (C/O) ratio $\ge$ 1. Recently, we proposed that these planets arise from beyond the snow line in circumstellar disks with oxygen abundances lower than those inferred in their parent stars (Madhusudhan et al. 2011b). In the solar system, the C/O ratio remains poorly constrained in the giant planets because obtaining a measurement of the water abundance below the meteorologically-active layer is difficult (Taylor et al. 2004). Data returned by the Galileo probe mass spectrometer in 1995 around the one-bar pressure level in Jupiter's atmosphere has provided carbon, nitrogen, sulfur, argon, krypton and xenon abundances (Wong et al. 2004; Mahaffy et al. 2000) that are relatively well matched by formation scenarios based on solar nebula models assuming solar elemental composition (Owen \& Encrenaz 2006; Mousis et al. 2009) -- what we refer to here as ``protosolar'' (see Table \ref{data} for the list of available constraints). Below expected water condensation level ($\sim$5 bar for solar abundance), the measured oxygen abundance was unexpectedly low, an effect typically attributed to the dynamics of the region within which the probe descended (Orton et al. 1998), but which we argue here could also partly reflect a bulk abundance lower than predicted by existing formation models. We point out that there is currently no observational evidence supporting the assertion that the oxygen abundance is strongly supersolar ($\ge$ 5 $\times$ solar) in Jupiter's interior. This statement results from models and theories proposed by different groups, including several authors of this work (Gautier et al. 2001; Alibert et al. 2005a,b; Hersant et al. 2004, 2008; Mousis et al. 2009). In fact, the water abundance has been sampled by the Galileo probe at only two different altitudes within the expected water condensation level in the Jovian atmosphere and the inferred O/H never exceeds 0.3 $\times$ solar even at $\sim$20 bar pressure (see e.g. Fig. 11 of Wong et al. 2004). These facts imply that the question of the bulk oxygen abundance in Jupiter should at least remain open.

The low water abundance measured by the Galileo probe has been previously hypothesized as due to Jupiter's formation from carbonaceous matter, such as tar, instead of icy planetesimals, requiring the formation of Jupiter inside the snow line (Lodders 2004). However, this hypothesis is not supported by the standard core-accretion model describing Jupiter's formation (Pollack et al. 1996; Alibert et al. 2005c). This model shows that Jupiter acquired the bulk of its mass in the cold outer region of the nebula, in which icy planetesimals are the dominant solids and precludes the idea that Jupiter might have accreted mainly from carbonaceous matter. { A low water abundance in the Jovian atmosphere could also be the natural outcome of the planet formation in a zone between H$_2$O and CO snowlines in the nebula ({\"O}berg et al. 2011). In this scenario, the giant planet's envelope is accreted from an oxygen-depleted gas from the nebula, as a result of the water condensation and incorporation at earlier epochs in the building blocks of the planetary core. However, this scenario predicts that the abundances of carbon, nitrogen and other ultravolatiles are solar in the envelope of Jupiter and does not explain the supersolar abundances measured by the Galileo probe.}

Here we find that all the observed elemental abundances of Jupiter can be explained consistently within the standard core-accretion model of Jupiter's formation beyond the snow line by only changing the C/O ratio in the formation zone.  The resulting O abundance in Jupiter's envelope then becomes moderately enriched compared to solar and is found to be consistent with values predicted by thermochemical models. To do so, we derived the elemental abundances in the envelope of Jupiter by tracking the chemical condensation and accretion of planetesimals through the planet's formation and evolution. We used a numerical model that relates the formation conditions of icy planetesimals accreted by Jupiter in the primitive nebula to the volatile abundances in its present atmosphere, the latter being determined from the amount of heavy elements accreted and dissolved in the planet's envelope during its growth (Gautier et al. 2001; Mousis et al. 2009; Madhusudhan et al. 2011b).

\section{Basic assumptions and modeling approach}

\begin{figure}
\begin{center}
\includegraphics[width=10cm,angle=0]{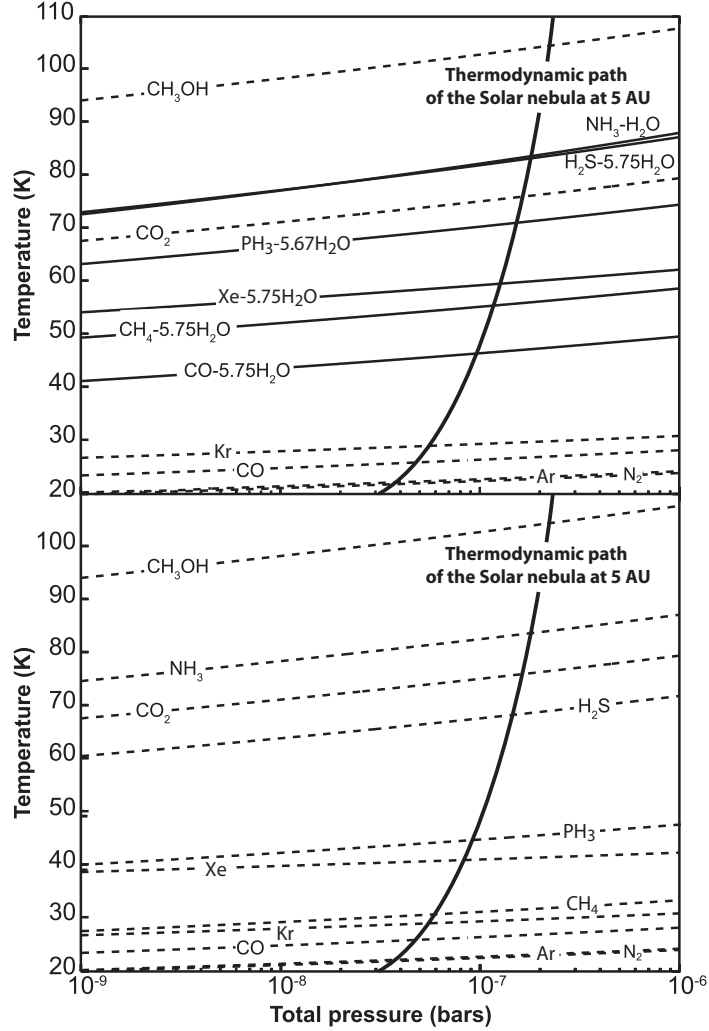}
\caption{Formation conditions of icy planetesimals in the solar nebula. Top panel: equilibrium curves of hydrate (NH$_3$-H$_2$O), clathrates (X-5.75H$_2$O or X-5.67H$_2$O) (solid lines), and pure condensates (dotted lines), and cooling curve of the solar nebula at 5 AU, assuming a full efficiency of clathration. Abundances of various elements are solar, with CO/CO$_2$/CH$_3$OH/CH$_4$ = 70/10/2/1, H$_2$S/H$_2$ = 0.5 $\times$ (S/H$_2$)$_\odot$, and N$_2$/NH$_3$ = 10 in the gas phase of the disk. Species remain in the gas phase above the equilibrium curves. Below, they are trapped as clathrates or simply condense. Bottom panel: same as top panel but with an oxygen abundance that is half the solar value. In this case, water does not exist in the disk and only pure condensates form.} 
\label{cool}
\end{center}
\end{figure}

Our model is based on a predefined initial gas phase composition in which all elemental abundances, except that of oxygen, reflect the bulk abundances of the Sun (Asplund et al. 2009) and describes the process by which volatiles are trapped in icy planetesimals formed in the protoplanetary disk. Oxygen, carbon, nitrogen, sulfur and phosphorus are postulated to exist only in the form of H$_2$O, CO, CO$_2$, CH$_3$OH, CH$_4$, N$_2$, NH$_3$, H$_2$S and PH$_3$. { We fix CO/CO$_2$/CH$_3$OH/CH$_4$~=~70/10/2/1 in the gas phase of the disk, a set of values consistent with the Interstellar Medium (ISM) measurements made by the Infrared Space Observatory (Ehrenfreund et al. 2000; Gibb et al. 2000) and at millimeter wavelengths from Earth (Frerking et al. 1982; Ohishi et al. 1992) considering the contributions of both gas and solid phases in the lines of sight. The dispersion of the ISM values is large and might reflect object-to-object variation as well as uncertainties of measurements but we stress that, among the possible molecular ratios, we selected those that are close to the cometary measurements (Bockel{\'e}e-Morvan et al. 2004).} Once the abundances of these molecules are fixed, the remaining oxygen gives the abundance of H$_2$O. Sulfur is assumed to exist in the form of H$_2$S, with an abundance fixed to half its protosolar value, and other refractory sulfide components (Pasek et al. 2005). We also consider N$_2$/NH$_3$ = 10/1 in the disk gas-phase, a value predicted by thermochemical models of the solar nebula (Lewis \& Prinn 1980). The process of volatile trapping in planetesimals formed in the feeding zone of proto-Jupiter, illustrated in Fig. \ref{cool}, is calculated using the equilibrium curves of hydrates, clathrates and pure condensates, and the thermodynamic path detailing the evolution of temperature and pressure at 5 AU (i.e. the current location of Jupiter) in the protoplanetary disk. We refer the reader to the works of Papaloizou \& Terquem (1999) and Alibert et al. (2005c) for a full description of the turbulent model of accretion disk used here. This model postulates that viscous heating is the predominant heating source, assuming that the outer parts of the disk are protected from solar irradiation by shadowing effect of the inner disk parts. In these conditions, temperature in the planet-forming region can decrease down to very low values (20 K; Mousis et al. 2009).

The top panel of Fig. \ref{cool} corresponds to the case where the gas phase abundances of various elements are solar, with the afore-mentioned gas phase molecular ratios. For each ice considered in this panel, the domain of stability is the region located below its corresponding equilibrium curve. The clathration process stops when no more crystalline water ice is available to trap the volatile species. The equilibrium curves of hydrates and clathrates derive from the compilation of published experimental work by Lunine \& Stevenson (1985), in which data are available at relatively low temperatures and pressures. On the other hand, the equilibrium curves of pure condensates used in our calculations derive from the compilation of laboratory data given in the CRC Handbook of Chemistry and Physics (Lide 2002). Note that, in the pressure conditions envisioned in the solar nebula, CO$_2$ is the only species that crystallizes at a higher temperature than its associated clathrate. We then assume that solid CO$_2$ is the only existing condensed form of CO$_2$ in this environment. In addition, we have considered only the formation of pure ice of CH$_3$OH in our calculations since, to our best knowledge, no experimental data concerning the equilibrium curve of its associated clathrate have been reported in the literature. In this case, the icy part of planetesimals is essentially made of a mix of pure condensates and clathrates. 

The bottom panel of Fig. \ref{cool} corresponds to the case of a disk composition similar to the one used in the top panel, except for the oxygen abundance that is set half the solar value. The subsolar O abundance adopted in the gas phase allows us to retrieve a composition of planetesimals that matches the value C/O = 1 in planetesimals formed in Jupiter's feeding zone (see Sec. 3). In this case, because the oxygen abundance is strongly depleted compared to previous case, this element is only distributed between carbon bearing species and the remaining water becomes zero in the initial gas phase of the protoplanetary disk. This implies that the icy part of planetesimals formed in such conditions in the protoplanetary disk is only made of pure condensates.

Finally, the intersection of the thermodynamic paths with the equilibrium curves of the different ices allows determination of the amount of volatiles that are condensed or trapped in clathrates at these locations in the disk following the approach depicted in Mousis et al. (2009) and Madhusudhan et al. (2011b). This method permits computation of the composition of the volatile phase p    resent in the planetesimals formed in Jupiter's feeding zone. The precise adjustment of the mass of these ices accreted by Jupiter and vaporized into its envelope allows us to reproduce the observed volatile enrichments. The fitting strategy is to match the maximum number of observed volatile enrichments and to determine the uncertainty range corresponding to this matching.

\section{Results}

\begin{figure}
\begin{center}
\includegraphics[width=14cm,angle=-90]{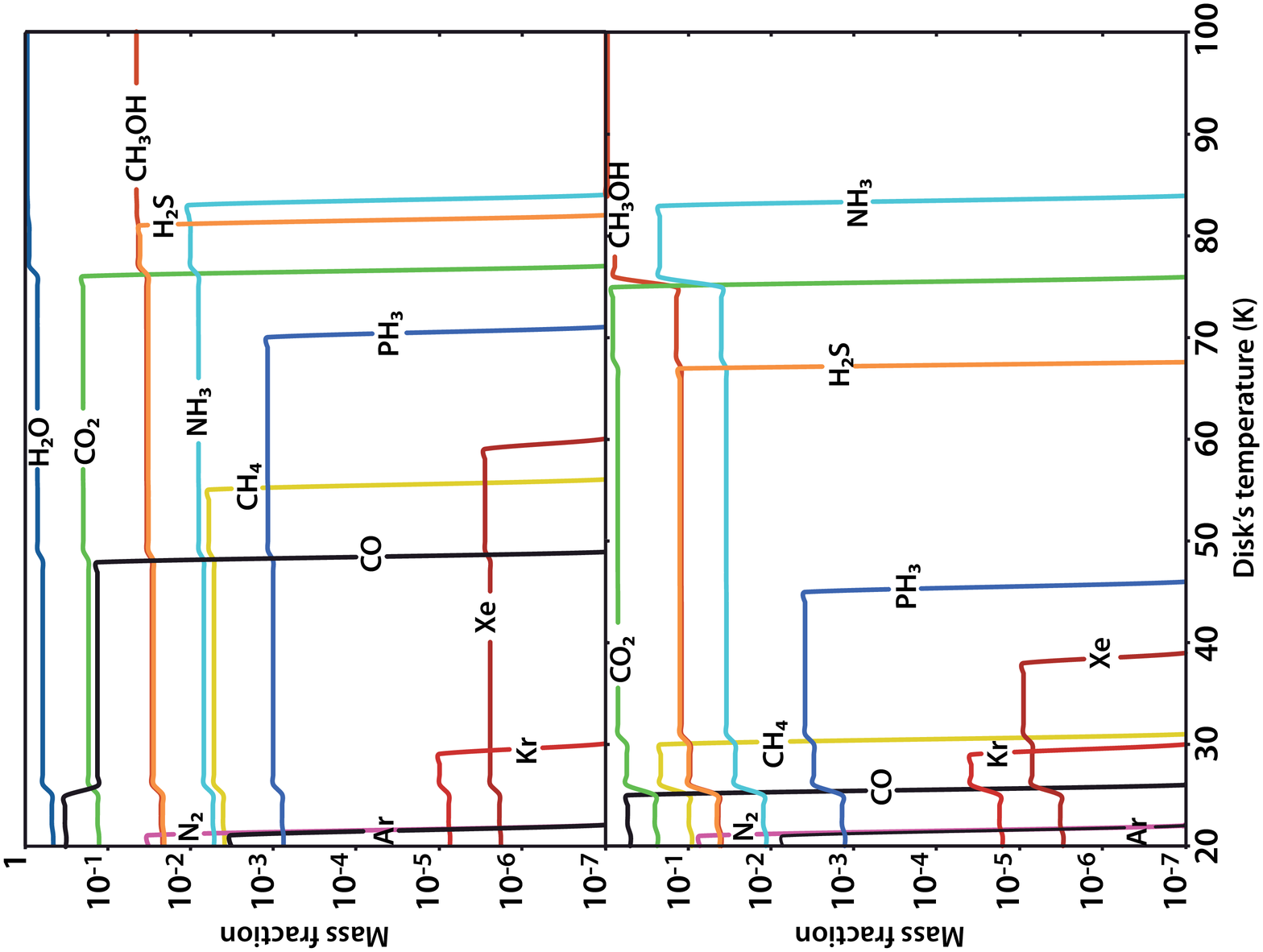}
\caption{Composition of the volatile phase incorporated in planetesimals as a function of their formation temperature at 5.2 AU in the solar nebula. Top panel: gas phase abundances of various elements are solar, with CO/CO$_2$/CH$_3$OH/CH$_4$ = 70/10/2/1, H$_2$S/H$_2$ = 0.5 x (S/H$_2$)$_\odot$, and N$_2$/NH$_3$ = 10 in the disk. Bottom panel: same as top panel but with an oxygen abundance that is half the solar value.} 
\label{comp}
\end{center}
\end{figure}

The formation conditions of these planetesimals, and thus their composition, strongly depend on the amount of crystalline water, i.e. the main oxygen-bearing molecule that is available in Jupiter's feeding zone. The presence of a high oxygen abundance in the nebula favors the formation of oxidizing molecules and also the trapping of volatiles in the form of hydrates at higher temperatures than those that would be expected for their pure condensates, which conversely are the relevant solid forms in the case of an oxygen-poor disk (Mousis et al. 2009; Madhusudhan et al. 2011b). Since fractionation occurs at the trapping/condensation epochs of the different volatiles (Mousis et al. 2009), the C/O ratio acquired by planetesimals differs from that in the gas phase. We thus conducted an iterative procedure allowing us to derive an oxygen abundance of $\sim$0.5 times its protosolar value in the nebula in order to get C/O = 1 in Jupiter's building blocks and envelope. Figure \ref{comp} represents the composition of ices agglomerated by planetesimals in the feeding zone of Jupiter as a function of their formation temperature and for two different compositions of the disk gas phase. We made the assumption that the planetesimals accreted by proto-Jupiter along its migration path, which is not expected to exceed 3--4 AU (Alibert et al. 2005b), formed from a homogeneous gas phase composition and that, at these locations, the disk cooled down to the same low temperature ($\sim$20 K) before its dissipation. These two conditions imply that the planetesimals accreted by the migrating planet shared a similar composition (Marboeuf et al. 2008; Mousis et al. 2009; Madhusudhan et al. 2011b) and that the matching of the observed elemental abundances does not depend on the time dependence of the gas giant's migration (Alibert et al. 2005b; Mousis et al. 2009). In the first case, corresponding to the canonical assumption that the composition of the disk is solar, water is the dominant volatile, irrespective of the formation temperature of planetesimals. In the second case, the gas phase composition of the disk is solar, except for the oxygen abundance that is set to half the protosolar value. 
The figure shows that carbon-bearing volatiles are the dominant volatile species trapped in planetesimals over the formation range considered here. Moreover, in this case, water does not exist in the formation zone of planetesimals since oxygen has been preferentially combined with C-bearing volatiles. { As a result, volatiles form pure condensates in the nebula at lower temperature than those usually encountered during their clathration when crystalline water is available.} Irrespective of the case considered, given the fact that the noble gas abundances have been measured supersolar in the atmosphere of Jupiter, this implies that these volatiles have been delivered in solid form to the giant planet (Mousis et al. 2009). In contrast, if noble gases remained in the gas phase of the disk, they would have been accreted in solar proportions in Jupiter's envelope (Madhusudhan et al. 2011b). The fact that noble gases have been delivered in solid form in Jupiter also implies that the formation temperature of planetesimals was as low as $\sim$20 K along the planet's migration path in the disk to allow their trapping or condensation.

\begin{figure}
\begin{center}
\includegraphics[width=14cm,angle=0]{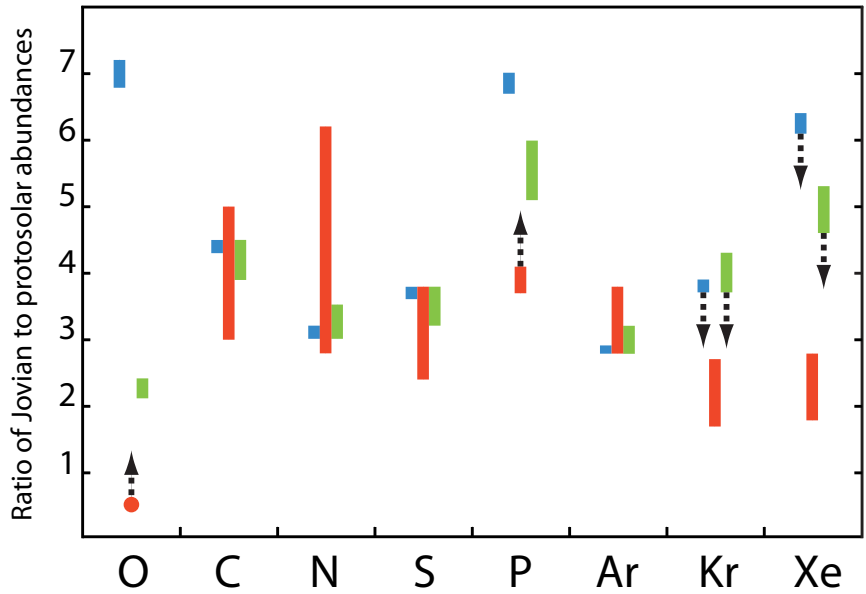}
\caption{Ratio of Jovian to protosolar abundances. Red bars and red dot correspond to observations. Green and blue bars correspond to calculations based on an oxygen abundance that is 0.5 and 1 times the protosolar value in the disk, giving C/O = 1 and 0.35 in Jupiter, respectively. The oxygen abundance is predicted to be 2.1--2.4 and 6.8--7.2 times protosolar in the cases of C/O~=~1 and 0.35 in Jupiter, respectively. Arrows up correspond to the possibility that the measured oxygen and phosphorus abundances are lower than their bulk abundances and arrows down to the possibility that planetesimals could be impoverished in krypton and xenon.} 
\label{fit}
\end{center}
\end{figure}

Once the composition of planetesimals has been calculated in the two cases, we adjusted the mass of heavy elements located in Jupiter's envelope to fit the maximum number of volatile abundances measured by the Galileo probe. Figure \ref{fit} represents the superimposition of the two fits with the measured volatile abundances and Table \ref{data} provides a summary of the different results. The figure shows that the same number of elements (carbon, nitrogen, sulfur and argon) is fitted in the two cases. However, the oxygen abundance predicted in Jupiter for an oxygen-depleted nebula is much closer to the measured abundance than the value predicted for a protosolar oxygen abundance. If the former case is correct, this supports the argument that the oxygen abundance in Jupiter derived from Galileo Probe water measurements reflects a bulk interior depletion of O relative to C, and is much less affected by atmospheric dynamical or meteorological processes than in the standard model. Neither calculation matches the observed phosphorus abundance, which is however only expected to provide lower bounds on the bulk abundance (Fletcher et al. 2009). The same remark applies for the observed krypton and xenon abundances but their relatively low values suggest the possibility of systematic error in their determination (Owen \& Encrenaz 2006).

\section{Discussion}

Our results, as discussed above, imply that a carbon-rich Jupiter provides a better explanation for the measured elemental abundances than the canonical case based on a protosolar oxygen abundance in the nebula. Our prediction of 2 $\times$ solar enhancement of oxygen in a carbon-rich Jupiter also agrees extremely well with recent constraints on the Jovian water abundance ($\sim$0.5--2.6 $\times$ solar) derived from tropospheric CO mixing ratios using thermochemical kinetics and diffusion models (Visscher \& Moses 2011). On the other hand, our model for the protosolar case predicts 7 $\times$ solar enhancement of oxygen in Jupiter which is ruled out by the thermochemical models (Visscher \& Moses 2011). { The important difference between the oxygen abundances in the two cases is a consequence of the presence or not of water ice in the giant planet's feeding zone. In the case of a solar oxygen abundance, water ice is the main O-bearing volatile present in the disk and accreted by Jupiter. The oxygen enhancement in the Jovian atmosphere is also amplified by the fact that, at the formation epoch of planetesimals, water condenses at much higher disk temperature and surface density compared to the other volatiles, thus increasing its mass fraction in solids. When the oxygen abundance becomes half solar in the nebula, the water abundance tends towards zero and the main O-bearing species supplied to the protoplanet atmosphere become CO and CO$_2$. These species condense at much lower disk surface density than water does and this effect increases the oxygen impoverishement in planetesimals accreted by proto-Jupiter.}

A carbon-rich Jupiter places stringent constraints on the formation conditions of planetesimals in the region of the solar system corresponding to the present asteroid belt and beyond. The presence of an oxygen-depleted zone located beyond the snow line in the nebula, which is needed to account for the presence of a carbon-rich Jupiter, may result from the inward evolution of the nebular snowline, which in some models leads to a water-depleted zone just inward of the condensation front (Cyr et al. 1998). In these models, the inward drift of icy particles, coupled to the diffusion of water vapor out past the snowline, tends to decrease the water gas phase abundance just inward of the snowline. Water would be the main volatile affected by this process since the other relatively abundant volatiles condense and decouple from gas just before the disk dissipation, which is expected to occur at $\sim$20 K in our model. The presence of such a zone, if restricted to a few AU, could still be consistent with the presence of water ice in the Jovian moons if their building blocks were formed at larger heliocentric distances in the nebula, or later when the snowline had moved inward and Jupiter was no longer in the water depleted zone. { In this case, the Jovian subnebula would be cold enough to allow formation of regular icy satellites from building blocks produced in the solar nebula (Canup \& Ward 2002; Mousis \& Gautier 2004; Alibert et al. 2005d). This idea} is supported by the recent measurement of the deuterium-to-hydrogen ratio in H$_2$O performed at Enceladus by the Cassini spacecraft showing that this satellite of Saturn was probably accreted from planetesimals similar in isotopic ratio to, and hence possibly condensed in the same formation region as Oort cloud comets ({ Horner et al. 2007}; Waite et al. 2009; Kavelaars et al. 2011). In this scenario, the composition of solids in the outer solar system would still remain dominated by water ice, except in the zone corresponding to the formation location of Jupiter. { Interestingly, it is still possible to argue that the Jovian regular icy satellites were accreted from building blocks condensed in a hot and dense subnebula fed by a CO-dominated gas coming from the solar nebula. In this case, CO would be converted to CH$_4$, making oxygen available for the formation of H$_2$O in the Jovian subnebula (Prinn \& Fegley 1981).}

A key observational test is the measurement of oxygen as water below the meteorological layer within Jupiter. A value of water about 2 $\times$ solar deep below the water clouds would confirm that Jupiter is carbon-rich. { The Microwave Radiometer aboard the recently launched Juno spacecraft will probe the deep atmosphere of Jupiter at radio wavelengths ranging from 1.3 cm to 50 cm to measure the planet's thermal emissions. This instrument will obtain measurements of water at pressures down to 100 bars deep in the Jovian atmosphere (Janssen et al. 2005), thereby constraining Jupiter's O/H and C/O ratios.}

\acknowledgements

We thank Jean-Marc Petit for valuable comments on the manuscript. O. M. is supported by CNES. JIL's contribution was supported by the Juno Project. TVJ work performed at the Jet Propulsion Laboratory, California Institute of Technology under a contract from NASA.  Government sponsorship acknowledged. NM acknowledges support from the Yale Center for Astronomy and Astrophysics through the YCAA postdoctoral Fellowship.

%
%

\clearpage

\begin{table*}
\centering \caption{Observed and calculated volatile enrichments relative to protosolar in Jupiter}
\begin{tabular}{lccc}
\hline \hline
Element 					& Constraints		&  Oxygen-depleted nebula	& Solar composition nebula 	\\
\hline
\multirow{2}{*}{O}			& 0.3--0.7$^a$		& \multirow{2}{*}{2.1--2.4}		& \multirow{2}{*}{6.8--7.2}		\\
						& 0.5--2.6$^b$		& 						&						\\
C     						& 3--5$^a$    		& 3.9--4.5   				& 4.3--4.5					\\
N       					& 2.8--6.2$^a$    	& 3.0--4.5   				& 3.0--3.2					\\
S						& 2.4--3.8$^a$		& 3.2--3.8					& 3.6--3.8					\\
P						& 3.7--4.1$^c$		& 5.1--6.0					& 6.7--7.0					\\
Ar              					& 2.8-3.8$^d$		& 2.8--3.2    				& 2.8--2.9					\\
Kr             					& 1.7--2.7$^d$		& 3.7--4.3    				& 3.7--3.9					\\
Xe              				& 1.8--2.8$^d$		& 4.6--5.3    				& 6.1--6.4					\\
\hline
\end{tabular}
\tablecomments{$^a$Wong et al. (2004); $^b$Visscher \& Moses (2011), $^c$Fletcher et al. (2009), $^d$Mahaffy et al. (2000)}
\label{data}
\end{table*}

\clearpage

\clearpage

\clearpage

\end{document}